\documentclass[prl, twocolumn, superscriptaddress]{revtex4-2}
\usepackage{amsmath,amssymb,bm}
\usepackage{hyperref}
\usepackage{graphicx}
\usepackage{epstopdf}
\usepackage{latexsym}
\usepackage{braket}
\usepackage{float}
\usepackage[normalem]{ulem}
\usepackage{comment}
\usepackage{mathtools}
\usepackage{array}
\usepackage{tabu}
\usepackage{multirow}
\usepackage[inline]{enumitem}
\usepackage{cleveref}
\usepackage[usenames,dvipsnames,table,xcdraw]{xcolor}
\usepackage[normalem]{ulem}
\def\BibTeX{{\rm B\kern-.05em{\sc i\kern-.025em b}\kern-.08em
    T\kern-.1667em\lower.7ex\hbox{E}\kern-.125emX}} 

\begin{document}
\title{Exclusion Statistics as a Thermodynamic Resource in Quantum Heat Engines}
\author{Sampurna Karmakar, Aziz Hasan and Sourin Das \\
{\emph{Department of Physical Sciences,
Indian Institute of Science Education and Research Kolkata \\
Mohanpur - 741246, West Bengal, India}}} 
\begin{abstract}
The maximum power extractable from a quantum thermoelectric heat engine operating with free fermion  carriers is bounded by the universal Whitney limit, $P_{\text{fermion}}^{\max} \simeq 0.0321\pi^2 k_B^2(T_L-T_R)^2/h$. We demonstrate that this bound is not fundamental to quantum heat engines but is instead an artifact of fermionic statistics. Within the nonlinear Landauer-B\"{u}ttiker framework, a bosonic working medium yields a strictly enhanced universal maximum power, $P_{\text{boson}}^{\max} = (\ln 2)^2\, k_B^2(T_L-T_R)^2/h$, exceeding the fermionic limit by a factor of $(\ln 2)^2/(0.0321\pi^2) \approx 1.52$. We propose magnon transport through a ferromagnetic spin chain as an experimentally viable bosonic realization. Incorporating Haldane fractional exclusion statistics with parameter $g$ provides a continuous interpolation between the bosonic ($g = 0$) and fermionic ($g = 1$) limits, revealing a monotonic enhancement of maximum power for $g < 1$ at reduced bias cost. These results establish quantum statistical exclusion as a previously unrecognized and independently tunable thermodynamic resource, opening performance regimes inaccessible to conventional carrier-engineering approaches.
\end{abstract}
\maketitle
\textit{\underline{Introduction}}: Heat engines convert thermal gradients into work by transporting energy between reservoirs at temperatures $T_L > T_R$, with efficiency bounded by the Carnot limit $\eta_{\mathrm{Carnot}} = 1 - T_R/T_L$, attainable only at vanishing power~\cite{callen1991thermodynamics}. The central problem of finite-time thermodynamics is to optimize the efficiency-power trade-off under nonequilibrium conditions~\cite{Curzon,Cis,Andresen,Esposito,Giuliano,Benenti,Ryabov,Long,Holubec,Seifert,Allahverdyan}. In quantum regime, this problem admits a microscopic formulation within the Landauer-B\"{u}ttiker framework~\cite{Landauer,Buttiker,Buttiker_1992}, where transport is governed by energy-resolved transmission functions and reservoir distributions, establishing a direct link between quantum transport and thermodynamic performance.

Thermoelectric devices provide a minimal realization of such quantum heat engines. A necessary condition for finite thermoelectricity and hence power generation is the breaking of particle-hole symmetry, since symmetric transport about the chemical potential yields exact cancellation between electron and hole contributions. Consequently, optimization has primarily focused on tailoring energy-dependent transmission functions that act as spectral filters. Within this framework, Whitney showed that, for free fermionic carriers, optimization over all admissible transmission functions yields a universal bound on efficiency at finite power, together with a corresponding bound on maximum power, hence defining a universal efficiency-power relation that serves as a benchmark for quantum thermoelectrics~\cite{Whitney_prl,Whitney_explain-prb}. These results rely on the assumption of fermionic statistics. 

Here, we demonstrate that this assumption is not merely a technical choice, but it fundamentally constrains thermoelectric performance. Within the nonlinear Landauer formalism, it is straightforward to extend towards the bosonic analogy, and we show that a bosonic working medium gives an elevated universal maximum power: $P^{\max}_{\mathrm{boson}} \simeq 1.52 \,P^{\max}_{\mathrm{fermion}}$ alongside enhanced efficiency-power characteristics. To connect this ideal bosonic bound with physical reality, we propose an experimentally viable setup using magnon transport across a driven ferromagnetic spin chain~\cite{Baigeng,Maekawa,Shankar,Brataas,Takei}, demonstrating that an intrinsic Zeeman gap can serve as the exact energy filter required to surpass the fermionic limit. Finally, to unify these two extreme regimes, we incorporate Haldane exclusion statistics~\cite{Haldane_exclusion}, characterized by a parameter $g$ that interpolates between bosonic occupancy ($g=0$) on one side and fractional occupancy ($g>1$) on the other, with the fermions ($g=1$) defining a boundary between the two via a generalized state counting principle. The resulting distributions~\cite{Wu} introduce an intrinsic particle-hole asymmetry~\cite{Chetan,karmakar2025}, providing a statistics-driven mechanism for thermoelectric optimization. Importantly, a physical realization of fractional statistics with $g=1/2$ emerges naturally via the spinon excitations in an antiferromagnetic Heisenberg spin chain~\cite{Haldane_exclusion,Murthy,BOUWKNEGT}, though a rigorous transport analysis of which lies outside the immediate focus of this study.

Our result reveals that thermoelectric performance is not solely determined by transmission engineering but is fundamentally shaped by carrier statistics. Haldane exclusion statistics thus provide a systematic route to surpass the bounds of fermionic transport, establishing quantum statistical exclusion as a thermodynamic resource that has not previously been characterized in the context of universal nonlinear power bounds. \\ \\
\textit{\underline{Universal bound of bosonic heat engine}}: Quantum heat engines can be modeled as a thermocouple circuit comprising two thermoelectrics with opposite responses, thereby an optimal single thermoelectric demonstrates how to optimize the thermocouple system~\cite{Whitney_explain-prb}. The trade-off between power and efficiency in quantum thermoelectrics was known to be bounded by Whitney limit for fermionic carriers~\cite{Whitney_prl}, establishing a universal benchmark for nanoscale transport. Whether this limit represents an absolute upper bound for all quantum heat engines remains an unexplored question. Here, we calculate the universal thermodynamic bound of a heat engine with a bosonic working medium operating within the nonlinear Landauer-B\"{u}ttiker scattering framework.

\begin{figure}[t]
    \centering
    \includegraphics[width=0.99\linewidth]{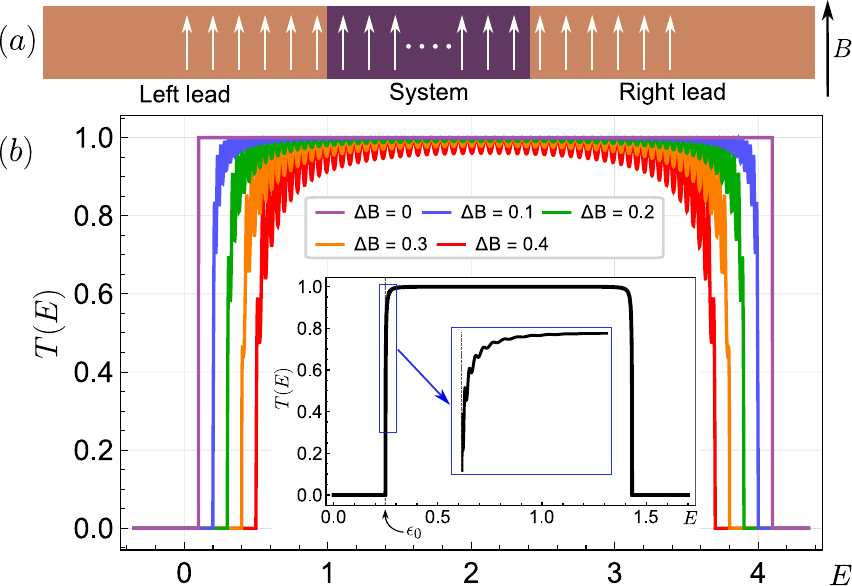}
    \caption{(a) Schematic of a ferromagnetic spin chain. (b) Transmission profile for a ferromagnetic spin chain with 99 spins and $B_{L/R}=B_S\pm \Delta B$. Here we have taken $B_S=0.1$ and $JS/g_l\mu_b =1$. The inset shows the transmission function used in the numerical analysis provided within a realistic regime.}
    \label{fig:transmission}
\end{figure}

Consider a thermoelectric system connected between two reservoirs, left (L) and right (R), which are maintained at temperatures $T_{L}$ and $T_{R}$ $( < T_L)$ and chemical potentials $\mu_{L}$ and $\mu_{R}$ $( >\mu_{L})$. Assuming the reservoirs are comprised of particles following Bose-Einstein distributions, their respective distribution functions read $f_B^{L/R}(E) = [e^{(E-\mu_{L/R})/k_B T_{L/R}} - 1]^{-1}$. Transport of particles through discrete channels of the system is governed by transmission probability $ \mathcal{T}(E)$. Following the nonlinear Landauer-B\"{u}ttiker framework, the charge and heat current flowing out of left/ right reservoir are~\cite{Hatano}
\begin{align}
    I_B^{L/R} =& \frac{q}{h}\int_{E(0)}^{\infty}dE\,\mathcal{T}(E)[f_{B}^{L/R}(E)-f_{B}^{R/L}(E)],\label{eq:charge_current}\\
    J_B^{L/R}=& \frac{1}{h}\int_{E(0)}^{\infty}dE(E-\mu_{L/R})\,\mathcal{T}(E)[f_{B}^{L/R}(E)-f_{B}^{R/L}(E)].\label{eq:heat_current}
\end{align}
where $E(0)$ marks the bottom of the conduction band and $q$ is particle charge. By summing the heat currents out of both reservoirs, we get the generated power:
\begin{align}
   P_B = \frac{\tilde{\mu}_BI_B^L}{q}= \frac{\tilde{\mu}_B}{h}\int_{E(0)}^{\infty}dE\,\mathcal{T}(E)[f_{B}^{L}(E)-f_{B}^{R}(E)],\label{eq:power}
\end{align}
where $\tilde{\mu}_B = \mu_R - \mu_L$ is the bias voltage. Note that for an energy-independent transmission function, $P_B\propto -\tilde{\mu}_B^2(<0)$, indicating the system dissipates energy rather than performing useful work. Therefore, achieving positive power generation strictly requires an asymmetric energy-dependent transmission function to break particle-hole symmetry. To maximize the power output over all physically admissible transmission functions $0 \le \mathcal{T}(E) \le 1$, we maximize the integrand of $P_B$, which is proportional to $f_B^L(E) - f_B^R(E)$. This yields a lower energy cutoff $\epsilon_0$ corresponding to the crossover energy where the two reservoir distributions intersect ($f_B^L(\epsilon_0) = f_B^R(\epsilon_0)$):
\begin{align}
    \epsilon_{0}=\mu_L+\frac{\Tilde{\mu}_B}{1-T_R/T_L},\label{eq:epsilon}
\end{align}
a threshold identical to that found for fermions~\cite{Whitney_prl}. Any non-zero transmission below this specific energy allows backward carrier flow, reducing the net power output. Consequently, the optimal transmission function that  maximizes power must take extremal values, leading to $\mathcal{T}_{\mathrm{opt}}(E)=\Theta(E-\epsilon_0)$, where $\Theta(E)$ is the Heaviside step function and the fact that $f_L-f_R$ decays fast enough as $E\rightarrow\infty$, ensures convergence of the integral. Finally, optimizing eq.~(\ref{eq:power}) with respect to $\Tilde{\mu}_B$ gives the maximum achievable power output and its corresponding optimized bias:
\begin{align}
    P_B^{\text{max}}&=(\ln2)^2\,k_B^2(T_L-T_R)^2/h,\label{Eq:Boson_power}\\
    \Tilde{\mu}_B^{\text{opt}}&=\ln2\, k_B(T_L-T_R).\label{Eq:Boson_cost}
\end{align}
Remarkably, this purely bosonic limit (eq.~(\ref{Eq:Boson_power})) substantially surpasses Whitney's universal fermionic bound, $ P_F^{\text{max}}\simeq 0.0321\pi^2\, k_B^2(T_L-T_R)^2/h$~\cite{Whitney_prl,Whitney_explain-prb}. Furthermore, this bosonic enhancement lowers the required bias cost (eq.~(\ref{Eq:Boson_cost})) compared to fermions ($\Tilde{\mu}_F=1.146\, k_B(T_L-T_R)$).\\ \\ 
\underline{\textit{Implementation in a physically realizable setup}}: To provide an experimentally realizable set up for our bosonic framework, we evaluate nonequilibrium magnon transport through a ferromagnetic spin chain~\cite{loss,Baigeng}. We consider the system as a 1D Heisenberg ferromagnet consisting of a finite central spin chain of $N$ sites coupled to two semi-infinite leads, subjected to longitudinal magnetic field bias (see fig.~\ref{fig:transmission} (a)). 
The central system (spanning sites 1 to $N$) features a nearest-neighbor ferromagnetic exchange interaction ($J > 0$) and a uniform local field $B_S$, while the left ($L$) and right ($R$) leads possess identical exchange terms alongside uniform magnetic fields $B_L$ and $B_R$, respectively. The total spin Hamiltonian is given by $\mathcal{H}_T=-J \sum_{i=-\infty}^{\infty} \mathbf{S}_i \cdot \mathbf{S}_{i+1} - g_l\mu_b\sum_{i=-\infty}^{\infty} B_i \,S_i^z,$ where $g_l$ is the Land\'e $g$-factor, $\mu_b$ is Bohr magneton and $B_i$ takes the values $B_L$, $B_S$ and $B_R$, depending on the respective region. In the low-temperature regime where spin fluctuations are small compared to the total spin magnitude ($\langle b_i^\dagger b_i \rangle \ll 2S$), a linearized Holstein-Primakoff transformation ($S_i^+\simeq\sqrt{2S}\, b_i, S_i^-\simeq\sqrt{2S}\, b_i^\dagger, \text{ and } S_i^z=S-b_i^\dagger b_i$)) maps the Hamiltonian onto an effective quadratic bosonic tight-binding model of magnons. The resulting Hamiltonians for the central system ($\mathcal{H}_s$), the left lead ($\mathcal{H}_L$), the right lead ($\mathcal{H}_R$), and the corresponding boundary coupling terms ($\mathcal{V}_L, \mathcal{V}_R$) read
\begin{align}
  &  \mathcal{H}_s= -J S \sum_{i=1}^{N-1} (b_ib_{i+1}^\dagger+h.c.)+(2 J S+g_l\mu_bB_S)\sum_{i=1}^N b_i^\dagger b_i,\nonumber\\
  &  \mathcal{H}_{L(R)}= -J S \sum_{i=-\infty(N+1)}^{-1(\infty)}(a_ia_{i+1}^\dagger+h.c.)\nonumber\\
    &\quad\quad\quad\quad\quad\quad+(2 J S+g_l\mu_bB_{L(R)}) \sum_{i=-\infty(N+1)}^{0(\infty)} a_i^\dagger a_i,\nonumber\\
     &   \mathcal{V}_{L}= -J S (a_{0}b_{1}^\dagger+h.c), \, \mathcal{V}_{R}= -J S (a_{N+1}b_{N}^\dagger+h.c).
\end{align}
where $a_i$ ($a_i^\dagger$) and $b_i$ ($b_i^\dagger$) denote the bosonic annihilation (creation) operators for the leads and the central system, respectively. We implement a nonequilibrium Green's function framework, using Lowdin partitioning to systematically integrate out the semi-infinite leads. This projects the reservoir boundaries onto the central system as energy-dependent self-energies: $\Sigma_{L/R}(E) = (J S)^2 g_{L/R}^s(E)$, where $g_{L/R}^s(E)$ is the surface Green's function of an uncoupled 1D semi-infinite lead. The retarded Green's function of the embedded central system is computed as $G^r(E) = \left[ (E + i\eta)I - H_{\text{eff}}(E) \right]^{-1}$, where $H_{\text{eff}}$ explicitly incorporates these boundary self-energies. Using the Caroli formula~\cite{Caroli_1971,meir,Jauho}, the magnon transmission function is evaluated as: $T(E) = \Gamma_L(E) \Gamma_R(E) |G^r_{1,N}(E)|^2$ where $\Gamma_{L/R}(E) = i[\Sigma_{L/R}(E) - \Sigma_{L/R}^\dagger(E)]$ represents the broadening functions at the contact sites and $G^r_{1,N}(E)$ is the retarded Green's function matrix element capturing the coherent propagation amplitude of a magnon across the length of the central chain.  In fig.~\ref{fig:transmission} (b), the transmission spectrum explicitly resolves both the filtering threshold imposed by the Zeeman terms and the spectral width governed by the exchange interactions. 
Within this setup, the magnetic fields applied to the leads dictate the underlying band edges while simultaneously serving as the driving bias. The transport properties are governed by magnons carrying pure spin current, following the same Landauer-B\"{u}ttiker formulations (eq.~(\ref{eq:charge_current}),(\ref{eq:heat_current})). The magnetic fields acting on the reservoirs establish effective magnetochemical potentials, $\mu_{\alpha} \equiv g_l\mu_b B_{\alpha}$ (with $\alpha = L,R$), which directly enter the Bose-Einstein distribution functions: $f_B^{\alpha}(E) = [e^{(E-g_l\mu_b B_{\alpha})/k_B T_{\alpha}} - 1]^{-1}$~\cite{Baigeng}. This establishes an exact thermodynamic mapping where the electrochemical potentials from our optimization framework are identified with the local magnetic field strengths. Consequently, the spatial difference in the longitudinal magnetic fields across the junctions provides the physical equivalent of a voltage bias, $\tilde{\mu} = g_l\mu_b(B_R - B_L)$, driving the magnon spin current across the central region.

To operate within the regime where the noninteracting magnon approximation remains strictly valid, the system must be maintained at temperatures low enough to suppress spin fluctuations, $k_B T \ll g_l\mu_B B$~\cite{loss}. For a baseline magnetic field of the order of $B_{L} \sim 2~\text{T}$, this constraint restricts the permissible operating temperature to the sub-Kelvin regime ($\sim 300~\text{mK}$). With these experimentally accessible parameters with reservoirs held at $T_{L} = 300~\text{mK}$ and $T_{R} = 10~\text{mK}$ and taking $J=3.5K\,k_B, S=1$, our proposed magnon heat engine demonstrates a substantial power output. Optimization of the magnetochemical bias yields $B_{R} = 2.149~\text{T}$, while the central region is set to $B_S = 2.15$ T to perfectly align with the threshold $\epsilon_0$ (see the inset of fig.~\ref{fig:transmission} (b)). This degree of field precision is routinely achieved in modern low-temperature magnon transport experiments~\cite{Czeschka}. Under these operational parameters, the realistic magnon spin chain generates a maximum power of $10.58$ pW. Remarkably, this performance captures over 91\% of the absolute universal bosonic bound ($11.58\text{ pW}$) derived from the ideal step function optimization, while fundamentally surpassing the conventional universal fermionic upper limit of  $7.64\text{ pW}$ under identical thermal gradients. This minor performance reduction relative to the absolute bound arises from the smooth, non-sharp lower edge of the intrinsic magnon transmission function, which deviates from a perfectly abrupt step profile. And, because the spin chain's bandwidth is larger than the characteristic energy scale where the distribution difference ($f_B^L - f_B^R$) drops to zero, the active transport window is structurally unconstrained by any upper band cutoff, preserving the near-ideal bosonic enhancement.\\ \\
\textit{\underline{Interpolating through Haldane's exclusion statistics}}:
Having established the performance benchmarks for the limiting cases of fermions and bosons, we turn to the continuous interpolation between them through Haldane exclusion statistics~\cite{Haldane_exclusion}.
\begin{figure}[t]
    \centering
    \includegraphics[width=0.96\linewidth]{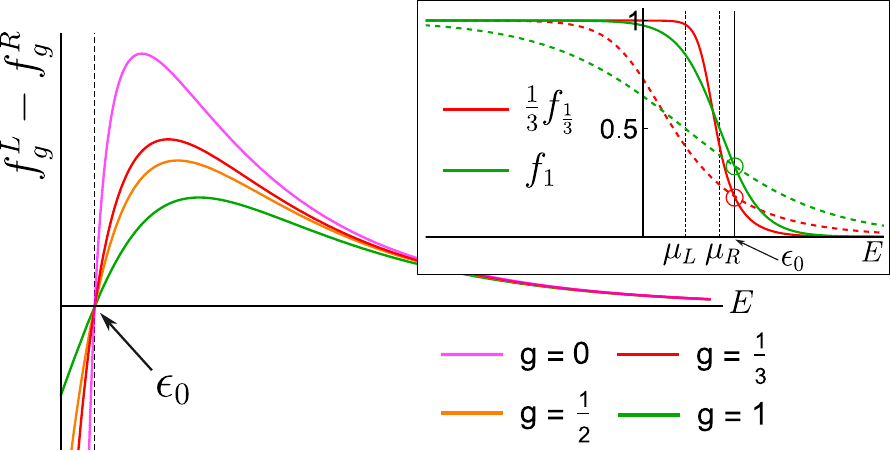}
    \caption{shows the lower limit ($\epsilon_0$) of the optimal transmission function at same temperature and voltage bias is identical for all the system of particles with different statistical parameter $g$. The inset shows the rescaled distribution function of the left (dashed lines) and right (solid lines) reservoirs intersects at same energy $\epsilon_0$, shown for $g=$ 1 and 1/3. $\mu_L$ and $\mu_R$ are shown with vertical dashed lines.}
    \label{fig:intersect}
\end{figure}
This states that for fixed system size and boundary conditions, the change in available single-particle states ($\Delta d$) and particle number ($\Delta N$) are related by $\Delta d=-g\,\Delta N$. Here the rational parameter $g$ interpolates between bosons ($g=0$) and fermions ($g=1$). For an ideal fractional gas, this generalized Pauli principle yields the distribution~\cite{Wu}: $f_g(E;\mu,T)=1/(\mathcal{W}(x,g)+g)$, where $x=(E-\mu)/(k_\text{B} T)$ (for particle energy $E$, temperature $T$, and chemical potential $\mu$), and $\mathcal{W}(x,g)$ satisfies: $\mathcal{W}(x,g)^g,[1+\mathcal{W}(x,g)]^{1-g}=e^{x}$. We aim to construct a heat engine operating with such fractional carriers to systematically track how statistical exclusion influences the bounds on power generation.

With a two-terminal setup similar to the previous section, we now consider that each reservoir is comprised of particles governed by the statistical parameter $g$. The particle current follows the same Landauer form (eq.~(\ref{eq:charge_current}),(\ref{eq:heat_current})) with the respective distribution functions $f_{g}^{L/R}(E)$. For an applied bias $\tilde{\mu}_{g}$, the generated power is $P_{g}=\tilde{\mu}_{g}\,I_{g}^{L}/q$. To determine the maximum achievable power, we fix the reservoir temperatures and optimize with respect to $\tilde{\mu}_{g}$ and all physically admissible transmission functions. Because the power integrand is proportional to $f_{g}^{L}(E)-f_{g}^{R}(E)$, only energy channels satisfying $f_{g}^{L}(E)>f_{g}^{R}(E)$ contribute positively to $P_g$. Since $f_g$ depends only on $(E-\mu)/T$, the intersection condition $f_g^L(\epsilon_0) = f_g^R(\epsilon_0)$ mathematically requires 
\begin{align}
    \epsilon_{0}=\mu_L+\frac{\Tilde{\mu}_g}{1-T_R/T_L},
\end{align}
independent of the value of $g$, as shown in fig.~\ref{fig:intersect}. 
\begin{figure}[t]
    \centering
    \includegraphics[width=0.95\linewidth]{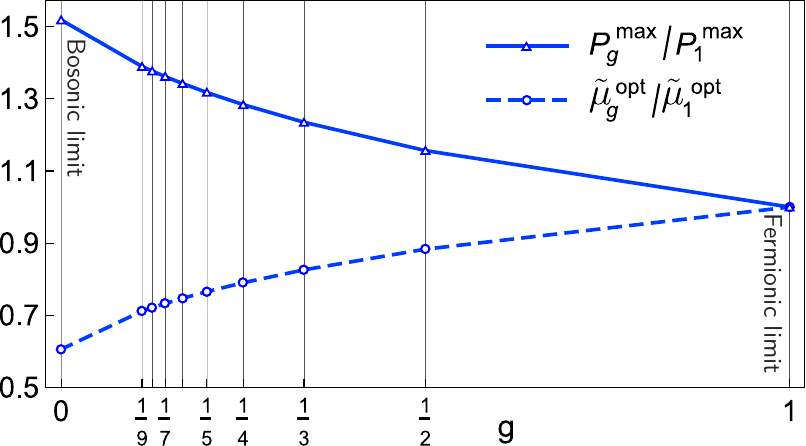}
    \caption{The solid line shows the ratio of maximum possible power output for heat engine with that of a fermion, as a function of the statistical parameter $g$. The dashed line shows the corresponding ratio of optimized biases.}
    \label{fig:Heat_power}
\end{figure}
Consequently, the optimal transmission function that maximizes power must take extremal values, leading to $\mathcal{T}_{\mathrm{opt}}(E)=\Theta(E-\epsilon_0)$. Next, we optimize $\Tilde{\mu}_g$ and get the maximum achievable power output and the corresponding bias as
\begin{align}
    P_g^{\text{max}}&=\xi(g)\, k_B^2(T_L-T_R)^2/h,\\
    \Tilde{\mu}_g&=\chi(g)k_B(T_L-T_R).
\end{align}
where $\xi(g)$ and $\chi(g)$ only depends on $g$. The interpolation between the maximum possible power for bosons and fermions is plotted as solid line in fig.~\ref{fig:Heat_power}. This shows for $g<1$, where we have the possibility of having enhanced power, the cost of bias is less than that of the fermion. This has been shown as dashed line in fig.~\ref{fig:Heat_power}. \\ \\
\textit{\underline{Maximum efficiency at finite power}}: The efficiency of a heat engine is defined as the ratio of the power output to the heat current extracted from the hot reservoir: $\eta_g = P_g/J_g^L$.
\begin{table}[b]
\caption{\label{tab:efficiency} Ratio of efficiency at maximum power $\eta_g(P_g^{\max})$ with Carnot efficiency as a function of exclusion statistics parameter $g$ for three different temperature ratios $T_R/T_L$.}
\begin{ruledtabular}
\begin{tabular}{cccc}
Statistical & \multicolumn{3}{c}{$\eta_g(P_g^{\max})/\eta_{\text{Carnot}}$} \\
\cline{2-4}
Parameter $g$ & $T_R/T_L = 0.05$ & $T_R/T_L = 0.2$ & $T_R/T_L = 0.4$ \\ 
\hline
$0$ (Bosons)   & $0.44$ & $0.407$ & $0.371$ \\
$1/3$          & $0.481$ & $0.448$ & $0.41$ \\
$1/2$ (Semions)& $0.49$ & $0.456$ & $0.418$ \\
$1$ (Fermions) & $0.504$ & $0.471$ & $0.433$ \\
\end{tabular}
\end{ruledtabular}
\end{table}
Since the maximum power operates at an efficiency $\eta_g(P_g^{\mathrm{max}})<\eta_{\mathrm{Carnot}}$ (see Table.~\ref{tab:efficiency}), a more practically relevant optimization problem is determining the maximum achievable efficiency for a specified finite power output $P_g \le P_g^{\mathrm{max}}$. 
\begin{figure}[t]
    \centering
    \includegraphics[width=0.95\linewidth]{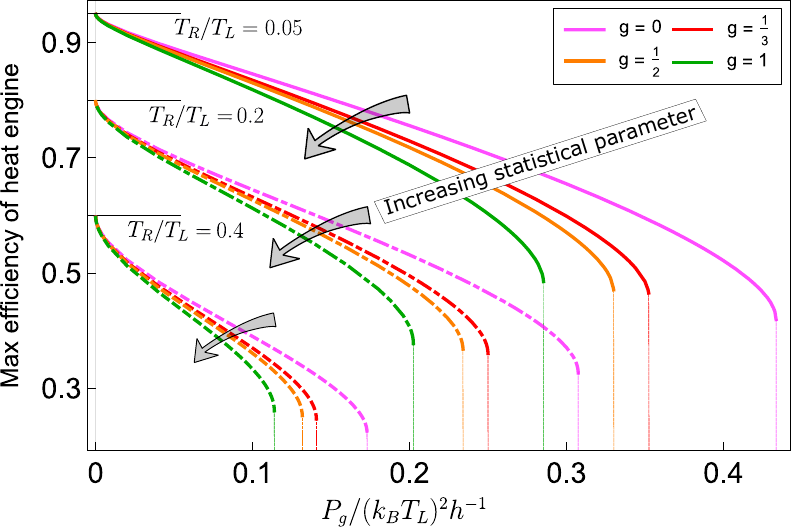}
    \caption{Efficiency vs power output for different statistical parameters with three temperature ratios. The vertical lines from each curve indicate the corresponding maximum power output. The horizontal black lines mark the Carnot efficiency for corresponding temperature ratios.}
    \label{fig:heat_eff}
\end{figure}
This constrained optimization over $\mathcal{T}(E)$ reveals that the optimal transmission function follows a ``boxcar"  profile~\cite{Whitney_prl,Whitney_explain-prb}: $\mathcal{T}_{\mathrm{opt}}(E)=\Theta(E-\epsilon_0)\,\Theta(\epsilon_1-E)$, where the lower bound $\epsilon_0$ remains defined by the intersection of the distribution functions, and $\epsilon_1$ is an adjustable upper energy cutoff. Operating at a target power requires a re-optimization of the parameters; specifically, for a given $g$ and a target power $P_g < P_g^{\mathrm{max}}$, we perform a systematic scan over the parameter space $(\tilde{\mu}_g, \epsilon_1)$. Because $\epsilon_0$ is bias-dependent, any change in $\tilde{\mu}_g$ shifts the entire transmission window. For each parameter set yielding the target power, we identify the maximum efficiency and by varying the target power from zero to $P_g^{\mathrm{max}}$, we obtain the efficiency-power characteristic curve as shown in fig.~\ref{fig:heat_eff}, serving as a benchmark for assessing the performance of any nanoscale thermoelectric medium. Physically, the introduction of $\epsilon_1$ filters out high-energy particles that contribute more to the heat current $J_g^L$ than the output power. As $P_g$ is reduced from $P_g^{\mathrm{max}}$, the optimal transmission window ($\epsilon_1 - \epsilon_0$) narrows. This suppresses the heat current more significantly than the power output, causing the efficiency to monotonically increase. Ultimately, the efficiency asymptotically approaches the Carnot limit for $\epsilon_1 \rightarrow \epsilon_0$, where the transmission window collapses to a delta function. Our results demonstrate that while all statistical species converge to $\eta_\text{Carnot}$ at zero power, their behavior at finite power is distinct. For any $P_g > 0$, bosons ($g=0$) exhibit a higher efficiency than fermions, indicating that the reduced cost of bias in the bosonic regime enables a more favorable power-efficiency trade-off.\\ \\
\textit{\underline{Conclusion}}: We have derived a universal bosonic bound $P_B^{\text{max}} = (\ln 2)^2 k_B^2(T_L-T_R)^2/h$, exceeding the fermionic bound by a factor of  approximately 1.52. We have shown that quantum statistical exclusion can be exploited as an independent thermodynamic resource to transcend the known performance limits in quantum transport. While prior spin-caloritronic studies utilize magnon-driven setups to optimize electronic charge and spin currents through interacting, Coulomb-blockaded quantum dots~\cite{Karwacki}, our approach shifts the focus entirely from local electronic parameters and interfacial conversion mechanics to the intrinsic statistics of the carrier medium itself. By grounding this paradigm in a realistic 1D ferromagnetic spin chain and generalizing it via Haldane fractional statistics, we demonstrate that tuning statistical exclusion provides a powerful, unexploited degree of freedom. This statistics-driven framework offers a novel avenue for quantum engine optimization, operating entirely parallel to conventional structural and transmission engineering.
\begin{acknowledgments}
The authors would like to thank Robert S. Whitney for valuable email communication. S.K. acknowledges support from the Prime Minister's Research Fellowship (PMRF) scheme of the Ministry of Education, Government of India (PMRF ID: 0501977). A.H. acknowledges University Grants Commission, India, for support in the form of a fellowship. S.D. would like to acknowledge the financial support from Anusandhan National Research Foundation (ANRF) under the MATRICS scheme (Grant No. [ANRF/ARGM/2025/002511/TS)]);  Ministry of Education, Government of India under the SPARC program (Project Code: [SPARC/2025-2026/P4086]) and National Quantum Mission under Quantum Algorithms Technical Group (TPN No.: 136428). The research of S.D. is supported partly by the International Center for Theoretical Sciences (ICTS), Bengaluru  through their Associateship Programmes. 
\end{acknowledgments}
\bibliography{citations}
\end{document}